\documentclass[twocolumn,prb,aps]{revtex4-2}
\usepackage{amsmath}
\usepackage{tensor}
\usepackage{physics}
\usepackage[pdftex]{graphicx}
\usepackage{subfigure}
\usepackage{float}
\usepackage{amssymb} 
\usepackage{xcolor}
\usepackage{hyperref}

\usepackage{etoolbox}
\apptocmd{\thebibliography}{\raggedright}{}{}
\def\sun{s}
\def\bub{b}
\def\loo{\ell}

\begin{document}
\author{
Enda Xiao$^1$ and
Chris A. Marianetti$^2$
}

\title{Anharmonic phonon behavior via irreducible derivatives: self-consistent perturbation theory and molecular dynamics}

\address{$^1$Department of Chemistry, Columbia University, New York, New York 10027, USA}
\address{$^2$Department of Applied Physics and Applied Mathematics, Columbia University, New York, New York 10027, USA}
\date{\today}

\begin{abstract}
Cubic phonon interactions are now regularly computed from first principles, and
the quartic interactions have begun to receive more attention. Given this
realistic anharmonic vibrational Hamiltonian, the classical phonon Green's
function can be precisely measured using molecular dynamics, which can then be
used to rigorously assess the range of validity for self-consistent
diagrammatic approaches in the classical limit.  Here we use the bundled
irreducible derivative approach to efficiently and precisely compute the cubic
and quartic phonon interactions of CaF$_2$, systematically obtaining the
vibrational Hamiltonian purely in terms of irreducible derivatives.  
non frequency shifts and linewidths, 
We demonstrate that the
4-phonon sunset diagram has an important contribution to the optical phonon
linewidths beyond $T=500$ K.  Reasonable results are obtained even at
$T=900$ K when performing self-consistency using the 4-phonon loop diagram and
evaluating the 3-phonon bubble and 4-phonon sunset diagrams post
self-consistency.  Further improvements are obtained by performing
quasiparticle perturbation theory, where both the 4-phonon loop and the real
part of the 3-phonon bubble are employed during self-consistency.  
Our irreducible derivative approach to self-consistent perturbation theory is a robust
tool for studying anharmonic phonons in both the quantum and classical regimes.
\end{abstract}

\maketitle

\section{Introduction}
Lattice anharmonicity is essential to the understanding of many physical
properties of solids, such as  thermal expansion, thermal conductivity, etc. \cite{Wallace19980486402126}. 
Therefore, computing phonon interactions from first principles is a critical
task.  Preliminary calculations of cubic phonon interactions from
density functional theory (DFT) began decades ago using finite displacements \cite{Wendel1978950},
and density functional perturbation theory (DFPT) computations eventually followed \cite{Gonze19951096,Debernardi19951819}.
Cubic phonon interactions have since been computed in a wide range of crystals using both 
DFPT and finite displacements \cite{Lindsay2019050902}.
Quartic phonon interactions were investigated
early on as well using DFT with finite displacements \cite{Vanderbilt19895657,Narasimhan19914541}. 
While quartic phonon interactions have been formulated at the level of DFPT \cite{Gonze19951096},
we are not aware of any explicit calculations. 
Given that the number of derivatives
increases drastically with the order, precisely and efficiently executing finite
displacement calculations of quartic phonon interactions is critical.
Recently, an approach for computing phonons and
phonon interactions based on irreducible derivatives was put forward,  which
maximally uses group theory to reduce computational cost, and the resulting
gain in efficiency can be converted into gains in accuracy
\cite{fu_group_2019}. Encoding anharmonicity in terms of irreducible
derivatives has many advantages, including allowing a straightforward
comparison between different methods.  Whenever possible, it is critical to
separately assess the quality of the vibrational Hamiltonian versus the
solution to the vibrational Hamiltonian, as the latter will contain errors from both aspects.
Here we solely focus on computing the interacting phonon Green's function of a
particular realistic vibrational Hamiltonian containing up to quartic phonon interactions.

The vibrational Hamiltonian poses a non-trivial many-boson problem, and
therefore it is challenging to assess the quality of the method being used to
solve the vibrational Hamiltonian.  Imaginary time Quantum Monte-Carlo (QMC)
techniques \cite{Ceperley1995279,Ceperley1999438} can be used to accurately compute thermodynamic observables for
relatively large systems, and various applications exist in the literature
using realistic tight-binding potentials \cite{Ramirez2006245202,Ramirez2008045210}. 
However, it is far more challenging to
obtain highly accurate solutions of dynamical quantities, such as the real time
phonon Green's function.  Imaginary time QMC results can be approximately
analytically continued to the real axis \cite{Jarrell1996133}, but these approaches are uncontrolled
and are best restricted to determining peak locations in the Green's function \cite{Sorkin2005214304}.
There are several established techniques which can be used to approximate
real time quantum correlation functions \cite{Perez2009184105,Althorpe2021155}, such as centroid molecular dynamics and ring
polymer molecular dynamics, but these approaches have various limitations \cite{Habershon2013387} and
have not yet been used to compute phonon linewidths in realistic
systems; though recent studies are beginning to make average comparisons via
thermal conductivity \cite{Luo2020194105}.  A practical
approach for obtaining the Green's function on the real axis is to use
diagrammatic perturbation theory, but the problem is assessing whether or not a
sufficient number of diagrams have been computed. A partial solution to this
problem is to revert to the classical limit where the vibrational Hamiltonian
can be solved accurately using molecular dynamics, and then a perturbative
solution can be rigorously assessed. Success implies that the diagrammatic approach is robust
in the classical regime, and will likely be sufficient in the quantum regime 
for sufficiently weak anharmonicity.

Using molecular dynamics to measure classical dynamical correlation functions
is well established in the context of empirical potentials \cite{wang_tight-binding_1990}. There have been a
small number of studies which use an anharmonic vibrational Hamiltonian based
upon first-principles calculations to measure dynamical correlation functions
within molecular dynamics \cite{glensk_phonon_2019,esfarjani_heat_2011,shiomi_thermal_2011,murakami_importance_2013,shiga_origin_2014,zhou_compressive_2019},
as is in the present study. Our molecular dynamics is based purely on irreducible
derivatives, which we refer to as irreducible derivative molecular dynamics
(IDMD), and we have developed methods to precisely compute the irreducible
derivatives from first principles \cite{fu_group_2019}, ensuring that we are working with a
realistic vibrational Hamiltonian. A key goal of this work is to use the IDMD
results to assess diagrammatic perturbation theory in the classical limit, and we are not aware
of comprehensive comparisons in the existing literature.

One of the more popular approximations to the interacting phonon problem is a
variational theory which uses a Gaussian ansatz for the density matrix,
which was originally pioneered by Hooton \cite{hooton_li_1955}.  This approach
can be viewed as the Hartree-Fock (HF) approximation for interacting phonons
\cite{jones_theoretical_2011}, given that the theory variationally determines
the optimum non-interacting bosonic reference system which minimizes the free
energy.  Naturally, Hartree-Fock for phonons also has a clear diagrammatic
interpretation, and therefore it is an integral approach for solving our
anharmonic Hamiltonian in this work.  There are several popular approaches
which implement this Hartree-Fock approximation for phonons, and it is useful to
contrast them with our own implementation.  
The self-consistent phonon (SCP) approach of Ref. \cite{tadano_quartic_2018},
later referred to as the first-order self-consistent phonon (SC1) approach \cite{tadano_first_2022},
uses compressive sensing \cite{zhou_lattice_2014,zhou_compressive_2019} to fit
the anharmonic vibrational Hamiltonian, and then the Hartree-Fock equations are solved
self-consistently 
in the usual manner.  Compressive sensing is useful given
that it can efficiciently fit the anharmonic terms to a relatively
small set of forces, but it is unclear how precisely it recovers 
the anharmonic terms as compared to the numerically exact answer, such as what can be
obtained with our irreducible derivative approach \cite{fu_group_2019}.
The SCP approach has been  used to compute
temperature dependent phonon dispersions \cite{tadano_first-principles_2018,tadano_first_2022,tadano_self-consistent_2015,tadano_quartic_2018,masuki_anharmonic_2022}
and  thermal expansion \cite{oba_first-principles_2019,masuki_ab_2022}.

Another approach for executing
Hartree-Fock is the stochastic self-consistent harmonic approximation (SSCHA)
\cite{errea_anharmonic_2014,bianco_second-order_2017}, which circumvents the need to Taylor series
expand the Born-Oppenheimer potential by performing a stochastic sampling of
the gradient of the trial free energy.  If executing the
Taylor series is prohibitive, potentially because very high orders are needed
to capture the relevant physics, a stochastic approach might be the only
practical method to execute Hartree-Fock. Another important aspect of the SSCHA is that the
full variational freedom of the Hartree-Fock ansatz is explored, allowing for the expectation values
of the nuclear displacements to be included as variational parameters. However, the SSCHA has its own
computational limitations for proper sampling,
 and the efficacy of the SSCHA is inherently tied to the
Hartree-Fock approximation.  
The SSCHA has been used in the computation of soft-mode driven phase transitions
\cite{errea_anharmonic_2011,bianco_second-order_2017}, charge density wave transitions
\cite{leroux_strong_2015}, and superconducting properties
\cite{errea_first-principles_2013,errea_anharmonic_2014,errea_high-pressure_2015,errea_quantum_2016,
borinaga_anharmonic_2016,borinaga_anharmonic_2016-1,bianco_high-pressure_2018,errea_quantum_2020,monacelli_black_2021}.
An earlier stochastic approach is the self-consistent \emph{ab initio} lattice-dynamical method (SCAILD) \cite{souvatzis_entropy_2008,souvatzis_self-consistent_2009}, which 
can be viewed as an approximation to the classical limit of the SSCHA.
The SCAILD approach has been used in the computation of temperature dependent phonon dispersion and structural phase transitions \cite{lazar_temperature-induced_2011,soderlind_high-temperature_2012,di_gennaro_role_2013},
and was later extended to the quantum case \cite{van_roekeghem_anomalous_2016,van_roekeghem_quantum_2021}; 
moving the method closer to the SSCHA. 
Another popular approach is the temperature-dependent effective potential
approach (TDEP) \cite{hellman_lattice_2011}, which uses the forces from a
classical \emph{ab initio} molecular dynamics trajectory as a source of data
to parameterize an effective quadratic potential. 
TDEP is not based on the Hartree-Fock approach: the intent is to faithfully recover the expectation value of the potential energy
of an \emph{ab initio} molecular dynamics simulation, and use the resulting effective quadratic potential to evaluate 
the corresponding harmonic quantum free energy \cite{hellman_temperature_2013}.
A recent application of the TDEP method changes course and adopts the stochastic sampling method
of the SSCHA \cite{heine_theory_2022}. 
It should be emphasized
that all of the aforementioned approaches could be directly applied to the anharmonic
Hamiltonian in the present study, but this would offer no benefit as compared
to our own Hartree-Fock solution. 
The application of SCP would only test
whether the compressive sensing approach faithfully reproduces our
irreducible derivatives, while the SSCHA would only test the
accuracy of the stochastic sampling. 
While Hartree-Fock is a key method that is used to study interacting phonon systems, it will
also be important to go beyond Hartree-Fock, motivating the evaluation of higher order diagrams.

\begin{figure}[htb]
\centering
\includegraphics[width=\linewidth, clip]{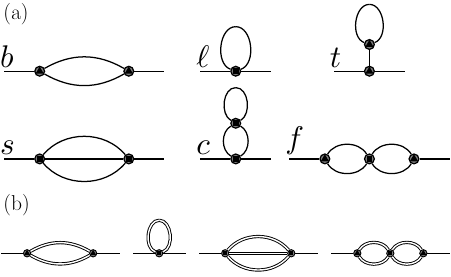}
\caption{(a) A schematic of the $O(\lambda^2)$ and $O(\lambda^4)$ bare self-energy diagrams considered in this work;
corresponding labels are provided. Vertices are bare cubic or quartic phonon interactions, while lines
are the non-interacting phonon Green's functions.
(b) A schematic of the dressed self-energy diagrams evaluated in this work.
Vertices are bare cubic or quartic phonon interactions, while lines
are self-consistent phonon Green's functions obtained using the Hartree-Fock or quasiparticle approach. 
}
\label{diagrams}
\end{figure}

Diagrammatic perturbation theory  is a
conventional method used to compute the phonon Green's function on the real axis \cite{Kokkedee1962374, maradudin_scattering_1962}. 
We consider all $O(\lambda^2)$ self-energy diagrams and selected $O(\lambda^4)$ diagrams \cite{tripathi_self_energy_1974} (see Figure \ref{diagrams} for schematics and labels). 
The colloquial diagram names of bubble, loop, tadpole,
sunset, cactus, and figure-eight are abbreviated as $\bub$, $\loo$, $t$, $\sun$, $c$, and $f$, respectively.
The mathematical definitions for the $\bub$, $\loo$, $\sun$, $c$, and $f$ diagrams are given in equations 
17, 12, 19, 21, and 24 of Reference \cite{tripathi_self_energy_1974}, respectively, and 
the $t$ diagram is defined in equation 2 of Reference \cite{lazzeri_anharmonic_2003}.
The classical limit of the $O(\lambda^2)$ diagrams (i.e. $\bub$, $\loo$, and
$t$) yield a linear temperature dependence for the self-energy, while the
classical limit of the $O(\lambda^4)$ diagrams
yield a quadratic temperature dependence.
It should be noted that the $t$ diagram is zero in the fluorite crystal structure \cite{maradudin_scattering_1962,lazzeri_anharmonic_2003}. 
The inclusion of all $O(\lambda^2)$ diagrams is clearly essential.
The imaginary part of the $\bub$ diagram is widely used in the context of perturbative thermal 
conductivity calculations \cite{broido_intrinsic_2007,Lindsay2018106}, and classically will provide
the only contribution to the linewidth to first order in temperature. 
The $\loo$ diagram is purely real,  and thus does not influence the phonon lifetime, 
and classically will provide a contribution to the phonon lineshift 
to first order in temperature. The real part of the $\bub$ and $\loo$ diagrams often oppose each other \cite{lazzeri_anharmonic_2003},
which justifies the success of using the bare phonon frequencies and the
scattering mechanism of the $\bub$ diagram in the linearized Boltzmann transport equation.
Our selection of $O(\lambda^4)$ diagrams will prove to be sufficient when used in conjunction with
self-consistent perturbation theory.
The importance of the $\sun$ diagram might be expected given that
it has a large influence on the phonon lifetime
at room temperature and beyond in select systems \cite{Kang2018575,Tian2018582,Li2018579}.
An important technical point in our work is that all diagrams are evaluated using the
tetrahedron method \cite{blochl_improved_1994}, which is important to efficiently achieving
convergence and removes issues associated with smearing parameters that are typically employed.

Including higher order diagrams beyond  $O(\lambda^4)$ is cumbersome, and a
more convenient approach is to perform a self-consistent perturbation theory in
terms of the dressed Green's function and skeleton diagrams \cite{Martin2016}.
Perhaps the simplest approach is the previously discussed Hartree-Fock
approximation for phonons. In the case where symmetry fixes the expectation
values of the nuclear positions, the Hartree-Fock approximation is simply a
self-consistent solution of the Dyson equation using the $\loo$ diagram. In
this case, an infinite number of bare diagrams would be summed, including the
$c$ diagram, though an infinite number of diagrams are still neglected, such as
$\sun$ or any diagram containing a cubic vertex.  The result of this approach
is a real, frequency independent self-energy which renormalizes the bare phonon
frequencies, though it does not produce a finite linewidth.  

One  approach for going
beyond Hartree-Fock is to use the resulting self-consistent Hartree-Fock Green's 
function to evaluate the $\bub$ diagram, which is known as the improved self-consistent (ISC) method \cite{goldman_improved_1968}, and the ISC is executed in the present work. 
The ISC based upon the SCP has been used in realistic computations of 
temperature dependent phonon spectrum \cite{tadano_self-consistent_2015,tadano_first-principles_2018},
soft-mode driven phase transitions \cite{tadano_first_2022,masuki_ab_2022}, 
thermal expansion \cite{oba_first-principles_2019,masuki_anharmonic_2022}, 
and thermal conductivity \cite{tadano_quartic_2018}. 
An approach which goes beyond the ISC is 
the  time-dependent self-consistent harmonic approximation
(TD-SCHA) \cite{monacelli_time-dependent_2021,lihm_gaussian_2021}, which constructs the Green's function
purely from quantities that are stochastically measured within Hartree-Fock. 
Recent applications of TD-SCHA within the bubble approximation \cite{monacelli_time-dependent_2021}, 
which is comparable to ISC, have produced reasonable predictions for the Raman and infrared
spectra in ice \cite{cherubini_microscopic_2021} and hydrogen under large pressures \cite{monacelli_black_2021},
in addition to providing a reasonable description of the inelastic x-ray scattering function
in NaCl and KCl at room temperature \cite{togo_lo-mode_2022}.
While the TD-SCHA goes beyond the ISC,
it is still a 
somewhat limited theory given that important diagrams are neglected. For example,
the TD-SCHA will not capture the effects of the sunset 
diagram, which is known to be critical even at room temperature in select systems \cite{Kang2018575,Tian2018582,Li2018579},
and we will demonstrate that it is important in CaF$_2$ beyond $T=500$ K.

An obvious approach for further potential improvement would be to use the
Hartree-Fock Green's function to evaluate both the $\bub$ and $\sun$
diagrams, which we execute in this study (see Figure S1 in Supplemental Material for a schematic \cite{supmat}).  
A final
approach is to use quasiparticle perturbation theory
\cite{Faleev_All_2004,Faleev_Quasiparticle_2006} and include the real
contribution of $\bub$ in the self-consistency and then use the resulting
Green's function to evaluate $\sun$, which we also execute.  We are not aware of
any previous quasiparticle perturbation theory calculations for phonons which
include frequency dependent diagrams, though very recently a ``one-shot"
approximation was performed \cite{tadano_first_2022}. A downside of
quasiparticle perturbation theory is that it is not a conserving approximation \cite{kutepov_electronic_2012}.

In this study we focus on the prototypical fluorite crystal CaF$_2$, where we
previously computed quadratic and cubic irreducible derivatives using DFT with
the strongly constrained and appropriately normed  functional \cite{xiao_validating_2022}.  
We demonstrated that the linewidth of the inelastic neutron scattering function
computed using the $\bub$ diagram was in good agreement with experiment throughout
the Brillouin zone at room temperature.
This stringent test validated the quality of
our quadratic and cubic irreducible derivatives, and therefore the underlying density functional
which was used to compute them, in addition to the exclusive use of the $\bub$
diagram to compute the Green's function. In the present work, we extend the
Taylor series to include quartic interactions, and our goal is scrutinize a
hierarchy of approximations which are used to compute the real and imaginary
part of the phonon self-energy.

\section{Methodology}

The vibrational Hamiltonian for CaF$_2$ is computed from density functional theory using the lone and bundled irreducible derivative
approaches (see Section \ref{sec:compdetails} for details).
Here we outline how to
compute the phonon self-energy \cite{maradudin_scattering_1962},
$\Sigma_{\boldsymbol{q} j j^{\prime}}(\omega)=\Delta_{\boldsymbol{q} j j^{\prime}}(\omega)-\textrm{i}\Gamma_{\boldsymbol{q} j j^{\prime}}(\omega)$,
in various approximations, which can be used to construct the
phonon lineshifts and linewidths. 
The contribution of a given self-energy diagram is denoted
$\Sigma_{\boldsymbol{q} j j^{\prime}}^{A,a}$, where
$a\in\{\bub,\loo,\sun,t,c,f\}$ labels a given diagram (see Fig. \ref{diagrams}) and
$A\in\{o,HF,QP\}$ labels which Green's function was used to evaluate the
diagram; where $o$, $HF$, and $QP$ correspond to the bare, Hartree-Fock, and
quasiparticle Green's function.
The $HF$ and $QP$ Green's functions are obtained by
self-consistently solving for the roots of $|\omega^{2}-\boldsymbol{V}_{\boldsymbol{q}}(\omega)|$,
where
\begin{align}
V_{\boldsymbol{q} j j^{\prime}}(\omega) &=
(\omega^0_{\boldsymbol{q} j})^{2} \delta_{j j^{\prime}}
+\left(2 \omega_{\boldsymbol{q} j}\right)^{\frac{1}{2}}
\left(2 \omega_{\boldsymbol{q} j^{\prime}}\right)^{\frac{1}{2}}
\Delta_{\boldsymbol{q} j j^{\prime}}(\omega),
\label{eq:sceq}
\end{align}
$\omega^0_{\boldsymbol{q} j}$ is the bare phonon frequency, and
$\omega_{\boldsymbol{q} j}$ is the renormalized phonon frequency.
In the case of $HF$, the functional form of $\Delta_{\boldsymbol{q} j j^{\prime}}$ is given
by the $\loo$ diagram, while for $QP$ the form of $\Delta_{\boldsymbol{q} j j^{\prime}}$ is given
by the combination of the $\loo$ diagram and the real part of the $\bub$ diagram.
The zeros of Eq. \ref{eq:sceq} deliver the updated renormalized frequencies and 
corresponding eigenvectors, which are then used to evaluate the updated $\Delta_{\boldsymbol{q} j j^{\prime}}$, and the
process is iterated until self-consistency is achieved. The resulting self-consistent $\Delta_{\boldsymbol{q} j j^{\prime}}$
is then used to construct the $HF$ or $QP$ Green's function.  For the  $QP$
case, we also test a ``one-shot" approximation, as recently implemented in
Ref. \cite{tadano_first_2022},  which replaces the
$QP$ self-consistency condition and instead uses
$\Delta^{HF,\loo}_{jj'}+\Delta^{HF,\bub}_{jj'}$ to construct the $QP$ Green's
function.

For a given scheme $A$, the self-energy is then
approximated $\Sigma_{\boldsymbol{q} j j^{\prime}}(\omega) \approx
\sum_{a}\Sigma_{\boldsymbol{q} j j^{\prime}}^{A,a} (\omega)$. Given that the
contribution from each diagram is additive, it can be useful to analyze
results for various combinations of diagrams. Therefore, we introduce a
notation to indicate which scheme and diagrams are used to construct a given
result as $\mathcal{S}^A_{ijk\dots}$, where $A$ labels the scheme and
$i,j,k,\dots$ indicate all diagrams evaluated. In this notation, 
the ISC approach \cite{goldman_improved_1968} is denoted $\mathcal{S}^{HF}_{\ell \bub}$, and the recent 
one-shot quasiparticle calculation in Ref. \cite{tadano_first_2022} is an approximation
to $\mathcal{S}^{QP}_{\ell\bub}$.
In this paper, the most diagrams evaluated in each scheme are
$\mathcal{S}^{o}_{\ell\bub\sun  fc}$, $\mathcal{S}^{HF}_{\ell\bub\sun f}$, and $\mathcal{S}^{QP}_{\ell\bub\sun f}$.
All of the diagrammatic approaches in this study are fully quantum mechanical approaches, but
we evaluate them in the classical limit  by replacing the 
Bose-Einstein distribution with $n(\omega) \rightarrow k_BT/\hbar \omega $ and
neglecting the zero-point contribution.

Standard molecular dynamics approaches can be used to obtain the classical 
solution of the anharmonic Hamiltonian constructed from irreducible derivatives, which we 
refer to as irreducible derivative molecular dynamics (IDMD). Using the IDMD trajectory,
the classical phonon spectral energy density $\mathcal{D}(\mathbf{q},\omega)$ at reciprocal point $\mathbf{q}$ 
is computed as 
\begin{align}
\nonumber 
\mathcal{D}(\mathbf{q},\omega) & =  \frac{1}{2 \pi N} \sum_{\mathbf{l}\mathbf{l}'} 
e^{-i\mathbf{q}\cdot (\mathbf{l} - \mathbf{l'}) }
\\ & \times
\sum_{dd'} 
\int d\tau e^{- i \omega \tau } \left< \mathbf{r}(\mathbf{l}d, \tau) \cdot \mathbf{r}(\mathbf{l}'d', 0) \right>,
\label{eq:classicalPSD}
\end{align}
where $\mathbf{l}$ labels the lattice translation, 
$d$ labels atoms within the primitive unit cell, 
$\mathbf{r}(\mathbf{l}d)$ is the  displacement associated with translation $\mathbf{l}$
and basis atom $d$, and
$N$ is the number of unit cells in the crystal.
The quantum $\mathcal{D}(\mathbf{q},\omega)$ can be constructed from the quantum single particle phonon Green's function
$D_{\boldsymbol{q}j}(\omega)$
as \cite{maradudin_scattering_1962, squires_introduction_2012}
\begin{align}
\label{eq:PSD}
 \mathcal{D}(\mathbf{q},\omega) =  
-\frac{\hbar n(\omega)}{2 \pi}  
\sum_j \frac{\Im(D_{\boldsymbol{q}j}(\omega))}{\omega_{\boldsymbol{q}j}^{0}}
\sum_{dd'}\frac{\mathbf{e}_{\mathbf{q}jd} \cdot \mathbf{e}_{-\mathbf{q}jd'}}
{\sqrt{M_d M_{d'}}},
\end{align}
where $M$ is the mass of the nuclei,
$n(\omega)$ is the Bose-Einstein distribution, and $\mathbf{e}_{\mathbf{q}jd}$ is the
polarization of atom $d$ in the mode $j$.
The imaginary part of $D(\mathbf{q},\omega)$ can be written in terms of the 
self-energy as \cite{maradudin_scattering_1962}
\begin{align}\nonumber
&\Im(D_{\boldsymbol{q}j}(\omega)))= \\
&\frac{-4 \big(\omega_{\boldsymbol{q}j}^{0}\big)^2 \Gamma_{\boldsymbol{q} j j}(\omega)}
{\big(\omega^{2}-(\omega_{\boldsymbol{q}j}^{0})^2-2\omega_{\boldsymbol{q}j}^{0}\Delta_{\boldsymbol{q} j j}(\omega)\big)^{2}
+\big(2\Gamma_{\boldsymbol{q} j j}(\omega) \omega_{\boldsymbol{q}j}^{0}\big)^2}.
\label{eq:D}
\end{align}
Equations \ref{eq:PSD} and \ref{eq:D} can be applied in the classical limit,
allowing one to relate the numerical measurements in Eq. \ref{eq:classicalPSD}
to the classical limit of the self-energy.
The simplest quasiparticle interpretation of some peak in $\mathcal{D}(\boldsymbol{q},\omega)$
which is identified with $\omega_{\boldsymbol{q} j}^0$ can be
characterized by the following trial function
\begin{align}\label{eq:fitfunc}
C_{0}
\frac{4\big(\omega_{\boldsymbol{q}j}^{0}\big)^2 C_{2}}
     {\big(\omega^{2}-(\omega_{\boldsymbol{q}j}^{0})^2-
     2\omega_{\boldsymbol{q}j}^{0}C_{1}\big)^{2}
     +\big(2 C_{2} \omega_{\boldsymbol{q}j}^{0}\big)^2},
\end{align}
which has three unknown coefficients
$C_{0}$, $C_{1}$, and $C_{2}$.
For a
given $\omega_{\boldsymbol{q} j}^0$, 
the corresponding peak in the IDMD measured  $\mathcal{D}(\boldsymbol{q},\omega)$
will be used to fit the three unknowns using linear regression. The energy window used
to determine which data are included
in the fit is 5 times the linewidth obtained from $\mathcal{S}^o_{\ell\bub\sun}$. In cases of overlapping peaks in $\mathcal{D}(\boldsymbol q,\omega)$, 
the corresponding peaks are individually resolved in the basis of the unperturbed eigenmodes and associated
with the corresponding $\omega_{\boldsymbol{q} j}^0$. 
The parameters resulting from the fitting process may be interpreted as the phonon lineshift
$\Delta_{\boldsymbol{q} j j}(\omega_{\boldsymbol{q} j}^0)=C_{1}$
and half linewidth $\Gamma_{\boldsymbol{q} j j}(\omega_{\boldsymbol{q} j}^0)=C_{2}$.

\section{Computational Details}
\label{sec:compdetails}

DFT calculations within the local density approximation (LDA)\cite{Perdew19815048} were performed using the projector
augmented wave (PAW) method \cite{blochl_improved_1994,Kresse19991758}, as implemented in the Vienna \textit{ab initio} simulation package (VASP) \cite{Kresse1993558,Kresse199414251,Kresse199615,Kresse199611169}. A plane wave basis with an energy cutoff of 600 eV was employed, along with a $k$-point density consistent with a centered k-point mesh of 20$\times$20$\times$20 in the primitive unit cell. All k-point integrations were done using the tetrahedron method with Blöchl corrections \cite{blochl_improved_1994}. The DFT energies were converged to within $10^{-6}$ eV, while ionic relaxations were converged to within $10^{-5}$
eV.  The structure was relaxed yielding a lattice parameter of 5.330 {\AA}, in agreement with previous work \cite{schmalzl_density-functional_2003}. The lattice constant is fixed in all calculations, and
we do not consider thermal expansion in the present work,
though it is straightforward to incorporate the strain dependence of 
the irreducible derivatives \cite{Mathis2022014314}.
The face-centered cubic lattice vectors are encoded in a $3\times3$ row stacked
matrix $\hat{\mathbf{a}}=\frac{a_o}{2}(\hat{\mathbf{J}}-\hat{\mathbf{1}})$,
where $\hat{\mathbf{1}}$ is the identity matrix and
$\hat{\mathbf{J}}$ is a matrix in which each element is 1.
The  quartic irreducible derivatives were calculated via the bundled irreducible
derivative (BID) approach  \cite{fu_group_2019}, and the quadratic and cubic terms were computed in
previous work \cite{xiao_validating_2022}. 
Up to 10 finite difference discretizations were evaluated for a given measurement, such that
robust error tails could be constructed and used to extrapolate to zero
discretization.  While the BID method only requires the absolute minimum number
of measurements as dictated by group theory, we tripled this minimum number in
order to reduce the possibility of contamination due to a defective
measurement. The LO-TO splitting was treated using the standard dipole-dipole
approach \cite{Gonze199710355,Giannozzi19917231} and implemented using irreducible
derivatives \cite{Mathis2022014314}.

\begin{figure}[htb]
\centering
\includegraphics[width=0.45\textwidth]{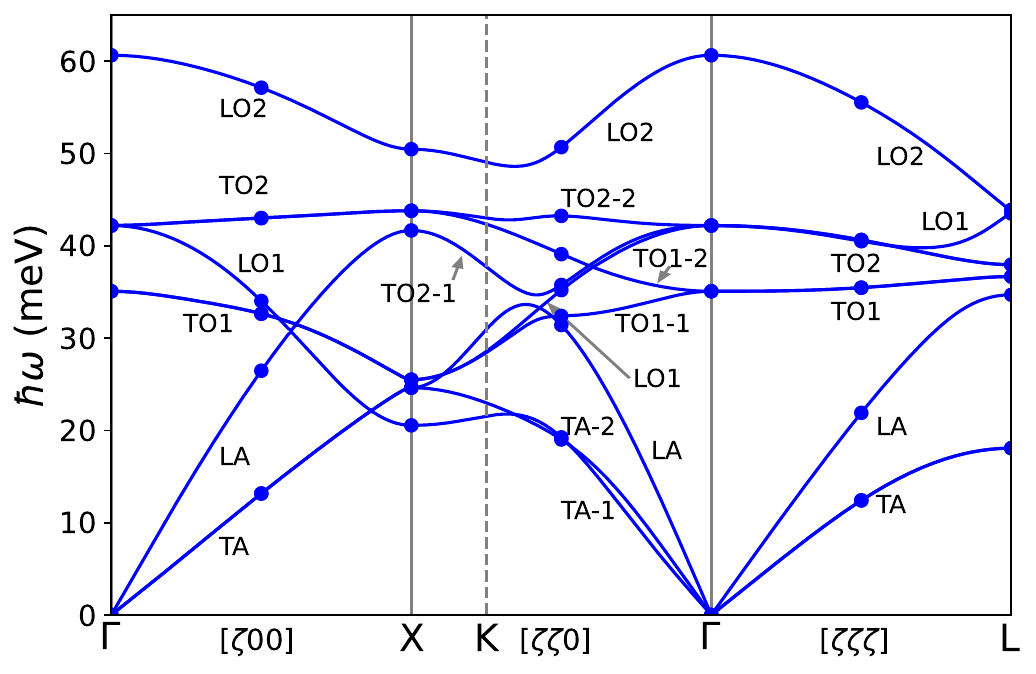}
\caption{The phonon dispersion of CaF$_2$, including branch labels. Points are computed from DFT and lines are a Fourier interpolation.
}
\label{dispersion}
\end{figure}

The Brillouin zone is discretized using a real space supercell
$\hat{\mathbf{S}}_{BZ}\hat{\mathbf{a}}$, where $\hat{\mathbf{S}}_{BZ}$ is an
invertible matrix of integers which produces superlattice vectors that satisfy
the point group \cite{fu_group_2019}. 
Two classes of supercells are used:
$n\hat{\mathbf1}$ 
and  $n\hat{\mathbf{S}}_{O}=n(4\hat{\mathbf{1}}-\hat{\mathbf{J}})$; where $n$ is a positive integer.
The second, third, and fourth order irreducible derivatives were computed for 
$\hat{\mathbf{S}}_{BZ}=4\hat{\mathbf1}$ (containing 64 primitive cells), $\hat{\mathbf{S}}_{BZ}=\hat{\mathbf{S}}_{O}$ (containing 16 primitive cells), and $2\hat{\mathbf1}$, respectively.
The quadratic and cubic irreducible derivatives have been previously computed and reported \cite{xiao_validating_2022},
and the quartic terms are reported in Supplemental Material \cite{supmat}.
A plot of the computed phonon band structure, including branch labels, is shown in Figure \ref{dispersion}.

The IDMD method is implemented using an interface to the
LAMMPS \cite{LAMMPS,plimpton_fast_1995} software package.  The irreducible
derivatives are Fourier interpolated to a  $10\times10\times10$ supercell.  The
Nose-Hoover thermostat \cite{nose_unified_1984} is used along with a 1 fs time
step.  For a given trajectory at each temperature, 30,000 steps are performed
for initialization followed by 600,000 steps. Five trajectories are performed
at each temperature, and all observables are averaged over the five
trajectories.  For all diagrammatic calculations, including self-consistent
calculations, irreducible derivatives are Fourier interpolated to a
$10\times10\times10$ supercell, and all integrations over the Brillouin Zone
involving the Dirac delta function are performed using the tetrahedron
method \cite{blochl_improved_1994}.  The real part of the self-energy was
obtained via a Kramers-Kronig transformation of the imaginary part.

\section{Results and Discussion}

\begin{figure}[htb]
\centering
\includegraphics[width=0.48\textwidth]{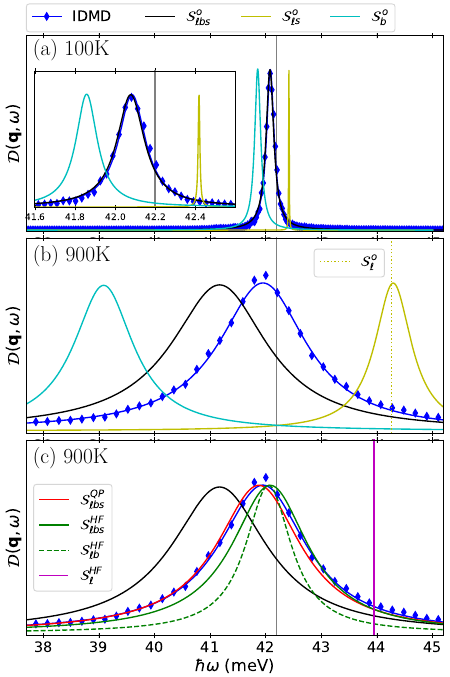}
\caption{Plots of $\mathcal{D}(\mathbf{q},\omega)$  at the $\Gamma$ point for an energy window around the  T$_{2g}$  modes.
The IDMD results are shown as blue diamonds, and the fit is shown as a blue curve.
The harmonic phonon frequency is denoted as a gray vertical line. 
($a$, $b$) Results at $T=100$ K and $T=900$ K for IDMD and bare perturbation
theory for various diagrams.  ($c$) Results at $T=900$ K for 
$\mathcal{S}^{QP}$, $\mathcal{S}^{HF}_{\loo\bub s}$, $\mathcal{S}^{HF}_{\loo\bub}$, and $\mathcal{S}^{HF}_{\loo}$, in addition to IDMD and $\mathcal{S}^{o}_{\loo\bub\sun}$ results.
Results for $\mathcal{S}^{o}_{\bub}$ and $\mathcal{S}^{o}_{\loo\sun}$ at 100K were rescaled by 0.94 and 0.06, respectively.  
Results for $\mathcal{S}^{o}_{\bub}$, $\mathcal{S}^{o}_{\loo\sun}$, $\mathcal{S}^{o}_{\loo\bub\sun}$, and $\mathcal{S}^{HF}_{\ell\bub\sun}$ at 900K were rescaled by 0.70, 0.43, 1.13, 0.90, and 0.54, respectively. 
}
\label{spectrum_o34}
\end{figure}

We begin by examining  $\mathcal{D}(\mathbf{q},\omega)$  at the
$\Gamma$-point in an energy window around the T$_{2g}$ modes (i.e. LO1 and TO2) to illustrate the various methods used to solve the vibrational
Hamiltonian, which contains up to quartic terms. We will 
explore a low temperature and a high temperature, though a very extensive survey is provided in 
Supplemental Material \cite{supmat}. We first consider the low temperature of $T=100$ K, where 
the classical perturbative approaches should be able to reasonably describe the
IDMD (see Figure \ref{spectrum_o34}, panel $a$). The IDMD
results are shown as blue diamonds, where each point is the result of binning all measurements
within a 0.02 meV window, and the blue line is the result of fitting equation \ref{eq:fitfunc} to the raw spectrum.
We begin by comparing  $\mathcal{S}^o_{\loo\bub\sun}$ to the IDMD spectra (see inset),
demonstrating excellent agreement with both the line shift and width, where the former is -0.12 meV
and the latter is 0.16 meV. It is interesting to further
decompose the result of $\mathcal{S}^o_{\ell\bub\sun}$ into
$\mathcal{S}^o_{\bub}$ and $\mathcal{S}^o_{\ell\sun}$, demonstrating that $\mathcal{S}^o_{\bub}$ is almost entirely responsible for the
linewidth, but it also substantially shifts the mode as well. However, the shift from $\mathcal{S}^o_{\bub}$ is partially cancelled by the shift from $\mathcal{S}^o_{\ell\sun}$, and it should be noted that the contribution from 
$\mathcal{S}^o_{\sun}$ is essentially negligible at this temperature. This compensation of the shift between $\mathcal{S}^o_{\bub}$
and $\mathcal{S}^o_{\loo}$ is not uncommon \cite{lazzeri_anharmonic_2003}, and it helps justify the success of thermal conductivity calculations solely
using the bare phonon frequencies and  
the $\mathcal{S}^o_{\bub}$ scattering mechanism when solving the linearized Boltzmann transport equation. The satisfactory performance of bare
perturbation theory implies that there is no need to consider self-consistent perturbation theory at this
temperature. 

We now proceed to the much higher temperature of $T=900K$, where the shift and
width are substantially larger (see Fig \ref{spectrum_o34}, panel $b$). As in
the low temperature case, $\mathcal{S}^o_{\bub}$ causes a downward shift and generates a
non-trivial linewidth. Unlike the low temperature case, $\mathcal{S}^o_{\ell\sun}$ not
only shifts the peak upward, but also generates a substantial linewidth. Given
that $\mathcal{S}^o_{\loo}$ is purely real, all of the linewidth contribution of
$\mathcal{S}^o_{\ell\sun}$ arises from $\mathcal{S}^o_{\sun}$, demonstrating that $\mathcal{S}^o_{\sun}$
can be an important contribution to the imaginary part of the self-energy at
higher temperatures.  Taking all contributions together, the
$\mathcal{S}^o_{\ell\bub\sun}$ provides a reasonable description of the IDMD result,
though there is a clear error in the shift, which suggests that self-consistent
perturbation theory may be needed. Indeed, $\mathcal{S}^{QP}_{\ell\bub\sun}$
provides an excellent description of the IDMD results (see Fig
\ref{spectrum_o34}, panel $c$). When moving down one level 
to $\mathcal{S}^{HF}_{\loo\bub\sun}$, the only notable degradation of the result is a small shift
to higher frequencies. It is also interesting to consider
$\mathcal{S}^{HF}_{\loo\bub}$ (i.e. the ISC) which shows a substantial
error in the linewidth. Furthermore, it is useful to consider the $f$ diagram, given that
it is recovered by TD-SCHA, and we find that it only has a very small contribution to both the 
linewidth and lineshift when evaluating any of the aforementioned schemes \cite{supmat}.

\begin{figure}[t]
\centering
\includegraphics[width=0.48\textwidth]{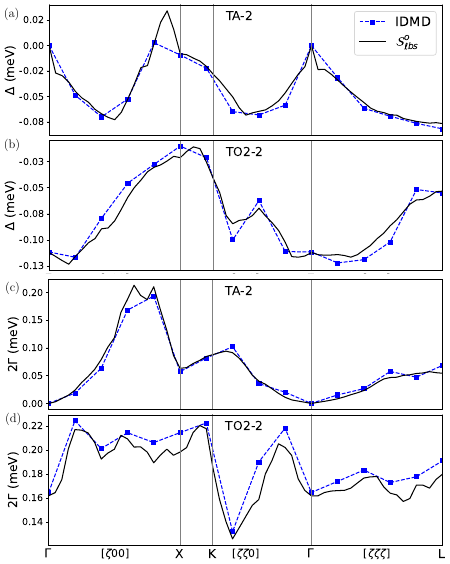}
\caption{ Phonon lineshifts ($a$, $b$) and linewidths ($c$, $d$) of the TA-2 ($a$, $c$) and TO2-2 ($b$, $d$) modes at
$T=100$ K along various paths through the Brillouin zone. IDMD and
$\mathcal{S}^o_{\loo\bub\sun}$ results are blue squares and black curves, respectively.
Curves between data points are guides to the eye.}
\label{o34_100}
\end{figure}

The preceding analysis carefully explored the results of different diagrams for a single mode, and 
we now proceed to survey select branches throughout the Brillouin zone. We begin by examining
the phonon line shift and width of the TA-2 and TO2-2 modes at $T=100$ K (see Figure \ref{o34_100}). 
For both the line shifts and widths, the $\mathcal{S}^o_{\ell\bub\sun}$ yields results that are close
to the IDMD. There are some regions where small differences can be noted, and care must be taken when
scrutinizing the results given the overall small magnitude of the numbers at hand. Nonetheless, the differences
mostly appear to arise from higher order diagrams, given that including self-consistency mostly tends to move
the diagrammatic solution closer to the IDMD (see Supplemental Material \cite{supmat}). The results for the remaining modes
are comparable \cite{supmat}.

\begin{figure}[t]
\centering
\includegraphics[width=0.48\textwidth]{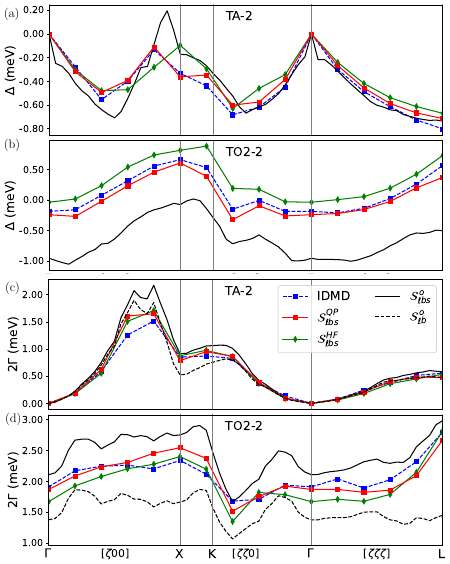}
\caption{ Phonon lineshifts ($a$, $b$) and linewidths ($c$, $d$) of the TA-2
($a$, $c$) and TO2-2 ($b$, $d$) modes at $T=900$ K along various paths through
the Brillouin zone.  IDMD, $\mathcal{S}^{QP}_{\ell\bub\sun}$,
$\mathcal{S}^{HF}_{\loo\bub\sun}$, 
$\mathcal{S}^o_{\loo\bub\sun}$,  and $\mathcal{S}^o_{\bub}$
results are blue squares, red curves, green curves, black curves, and dashed black curves,
respectively.  Curves between data points are guides to the eye.}
\label{o34_900}
\end{figure}

Having established the fidelity of our diagrammatic approaches and IDMD at low
temperatures, we now proceed to the high temperature of $T=900$ K where IDMD
can be used to judge the accuracy of different levels of bare and
self-consistent perturbation theory (see Figure \ref{o34_900}). As done
previously, we explore the TA-2 and TO2-2 modes, but here we consider
self-consistent perturbation theory as well given the deficiency of bare
perturbation theory at this temperature. We begin by examining the shift of the
TA-2 mode (panel $a$), where $\mathcal{S}^o_{\ell\bub\sun}$ yields good results in
certain portions of the zone, but performs poorly in selected regions.
Including self-consistency via $\mathcal{S}^{QP}_{\ell\bub\sun}$ and
$\mathcal{S}^{HF}_{\loo\bub\sun}$ tends to correct large deviations that are observed in
$\mathcal{S}^o_{\ell\bub\sun}$.  For the shift of the TO2-2 modes (panel $b$), the
$\mathcal{S}^o_{\ell\bub\sun}$  result yields a more systematic error, where the
results are shifted in a nearly uniform manner, and the result even has the
wrong sign in some cases. In this case, $\mathcal{S}^{HF}_{\loo\bub\sun}$ offers a
drastic improvement, and $\mathcal{S}^{QP}_{\ell\bub\sun}$ pushes the result even
closer to IDMD. The linewidths have similar behavior to the lineshifts in the
two respective cases (see panels $c$ and $d$). In the case of TA-2, it is
interesting to consider $\mathcal{S}^o_{\bub}$, given its importance in the context of
common treatments of thermal conductivity, and it is clear that it yields remarkable
results.  However, $\mathcal{S}^o_{\bub}$ performs poorly for the TO2-2 modes.
It is also interesting to consider the effect of the $f$ diagram, and it is shown to 
only have a small effect on both the linewidths and lineshifts \cite{supmat}.
The general trends that we have outlined here can also be seen in other modes,
and comprehensive results for all branches at temperatures of 100K, 300K, 500K, 700K, and 900K are included in Supplemental Material \cite{supmat}. 
We also provide a corresponding analysis where
all cubic interactions are set to zero \cite{supmat}.

\section{Conclusion}
In summary, we have computed the irreducible derivatives of CaF$_2$ up to fourth
order, defining the vibrational Hamiltonian. We use molecular dynamics to solve
the vibrational Hamiltonian, which we refer to as irreducible derivative
molecular dynamics (IDMD), yielding the real and imaginary part of the phonon
self-energy in the classical limit. The IDMD result was then used as a
benchmark for various levels of diagrammatic perturbation theory. At the low
temperature of $T=100$ K, we show that bare perturbation performs well using
the $\bub$, $\loo$, and $\sun$ diagrams (i.e. $\mathcal{S}^o_{\ell\bub\sun}$). While the
linewidth is reasonably well described by the $\bub$ diagram alone at $T=100$ K, the  $\loo$ diagram is also
necessary to properly capture the lineshift.  At the high temperature of
$T=900$ K, bare perturbation theory only performs well for the linewidths of
the acoustic modes, where even the $\bub$ diagram alone yields reasonable results.  Treating
the $\loo$ diagram self-consistently and evaluating the $\bub$ and $\sun$ diagrams post
self-consistency (i.e. $\mathcal{S}^{HF}_{\loo\bub\sun}$) is critical to obtaining
accurate lineshifts for all branches, in addition to obtaining accurate
linewidths for the optical modes.  Further improvement is normally obtained
when performing quasiparticle self-consistent perturbation theory, where the
real part of the $\bub$ diagram is used during self-consistency (i.e.
$\mathcal{S}^{QP}_{\ell\bub\sun}$). While we only executed self-consistent
perturbation theory in the classical limit in this work, it should be emphasized
that the quantum case has the same computational cost and poses no difficulty
beyond the classical case. 

The procedure outlined in this paper of assessing various levels of
diagrammatic perturbation theory in the classical limit using molecular
dynamics should be viable on nearly any system where the Taylor series can be
constructed. Once some class of diagrams is validated classically, it is likely
that the quantum counterparts will also perform well at lower temperatures in
the quantum regime so long as the anharmonicity is sufficiently weak.
Furthermore, the prescribed diagrams may be combined with other diagrammatic
approaches to scattering, such as defects or electron-phonon coupling in the
case of metals. An obvious application of the philosophy of this paper would be
phonon mediated thermal conductivity, where the molecular dynamics solution would yield
the thermal conductivity within Kubo-Green linear response, and various 
levels of self-consistent diagrammatic perturbation theory could be tested in conjunction
with the linearized Boltzmann transport equation in the classical limit.

\section{Acknowledgments}
We are grateful to Zhengqian Cheng for fruitful discussions.
This work was supported by the grant DE-SC0016507 funded by the U.S. Department of Energy, Office of Science.
This research used resources of the National Energy Research Scientific Computing Center, a DOE Office of Science User Facility supported by the Office of Science of the U.S. Department of Energy under Contract No. DE-AC02-05CH11231.

\clearpage
\bibliography{main}
\end{document}